\documentclass[amsmath,superscriptaddress,amssymb]{revtex4}
\def\beq{\begin{eqnarray}}
\def\eeq{\end{eqnarray}}
\def\kB{k_{\rm B}}
\def\lP{\ell_{\rm P}}
\def\lC{\ell_{\rm C}}
\def\tP{t_{\rm P}}
\def\mP{m_{\rm P}}
\def\Hz{{\rm Hz}}
\def\K{{\rm K}}
\def\gw{{\rm gw}}
\def\d{{\rm d}}
\def\s{{\rm s}}
\def\m{{\rm m}}
\def\g{{\rm g}}
\def\at{{\rm at}}

\begin{document}

\title{Quantum decoherence and gravitational waves}

\author{M.T. Jaekel$^*$}
\address{Laboratoire de Physique Th\'eorique de l'Ecole Normale Sup\'erieure, CNRS, UPMC\\
24 rue Lhomond, 75231 Paris Cedex 05, France\\
$^*$E-mail: jaekel@lpt.ens.fr}

\author{B. Lamine, A. Lambrecht and S. Reynaud$^{**}$}
\address{Laboratoire Kastler Brossel, Universit\'e Pierre et Marie Curie, CNRS, ENS\\
4 Place Jussieu, 75252 Paris Cedex 05, France\\
$^{**}$E-mail: reynaud@spectro.jussieu.fr}

\author{P. Maia Neto}
\address{Instituto de F\'{\i}sica, UFRJ, \\
CP 68528, 21945-970 Rio de Janeiro, Brazil}

\begin{abstract}
The quite different behaviors exhibited by microscopic and macroscopic systems with respect
to quantum interferences suggest the existence of a borderline beyond which quantum systems 
loose their coherences and can be described classically. 
Gravitational waves, generated within our galaxy
or during the cosmic expansion, constitute a universal environment
susceptible to lead to such a quantum decoherence mechanism. We assess this idea by
studying the quantum decoherence due to gravitational waves
on typical microscopic and macoscopic systems, namely an atom interferometer
(HYPER) and the Earth-Moon system. We show that quantum interferences remain 
unaffected in the former case and that they disappear extremely rapidly in the latter case.
 We obtain the relevant parameters which, besides the ratio of the system's mass to Planck mass,
 characterize the loss of quantum coherences. 
\end{abstract}

\keywords{quantum decoherence, gravitational waves}
\maketitle

\section{Introduction}

Quantum decoherence is a universal phenomenon which affects  all physical systems as soon as they
are coupled to a fluctuating environment. 
This effect plays an important role in the transition between quantum 
and classical behaviors, by washing out quantum coherences and thus justifying
a purely classical description 
\cite{Zeh70,Zurek81,Caldeira83,Joos85,Bal06}. 
This implies that quantum decoherence should be very efficient for 
macroscopic systems, while remaining inefficient for microscopic ones.
The quantum/classical transition would then introduce a  borderline between microscopic
and macroscopic systems. 

Existing experimental observations of quantum decoherence confirm these 
intuitions. 
Decoherence has only been seen on `mesoscopic' systems for 
which the decoherence time is neither too long nor too short, such as
microwave photons stored in a high-Q cavity \cite{Brune96} 
or trapped ions \cite{Myatt00}.
In such model systems, the environmental fluctuations are particularly
well mastered and the quantum/classical transition has been shown 
to fit the predictions of decoherence theory \cite{Raimond01,Raimond06}.

It has also been early remarked that Planck mass, that is the mass scale which can be 
built up on Planck constant $\hbar$, light velocity $c$ and Newton gravitation $G$, lies  
at the borderline between microscopic and macroscopic masses
\beq
&& \mP = \sqrt{\frac{\hbar c}{G}} \sim 22 \mu \g
\eeq
That is to say, one may define microscopic and macroscopic values of a mass $m$ 
 by comparing the associated Compton length $\lC$ to the Planck 
length $\lP$ 
\beq
m \lessgtr \mP &\Leftrightarrow & \lP = \sqrt{\hbar G\over c^3} \lessgtr \lC = \frac{\hbar }{mc} 
\eeq
It is tempting to consider that this property is not just an 
accidental coincidence but rather reveals a general consequence of fundamental
gravitational fluctuations \cite{Feynman,Karolyhazy66,Diosi89,Penrose96}.
Then, one is led to study the role that the fluctuating gravitational environment
might play in the transition from quantum to classical behaviors.

Here, we briefly discuss the quantum decoherence due to our local gravitational environment,
 namely the stochastic background of gravitational waves surrounding the Earth. 
Details can be found in previously published work \cite{Reynaud01,Reynaud02,Lamine02,Reynaud04,Lamine06}. 
First, taking the example of the atomic interferometer HYPER,
we show that gravitational waves do not lead to a significant decoherence 
at the microscopic level. We then show that, on the contrary, scattering of gravitational waves
 is the dominant decoherence mechanism, and an extremely efficient one, for macroscopic systems such as the 
Moon around the Earth. We also go beyond the simple scaling arguments just given above
by providing estimates of gravitational quantum decoherence
depending not only on the mass of the system, but also on its velocity,
on its geometry and on the noise spectrum
characterizing the gravitational fluctuations.

\section{Gravitational environment}	

We first describe the fundamental fluctuations of space-time  which originate from
our gravitational environment and which are bound to 
play a crucial role in quantum decoherence.
For current quantum systems which are only sensitive to frequencies lying far below Planck 
frequency, general relativity provides the appropriate description of 
gravitational phenomena \cite{Jaekel95}, even if it may ultimately be replaced by a 
theory of quantum gravity. 
It follows that the relevant spacetime fluctuations which constitute
our gravitational environment are simply the gravitational waves 
predicted by the linearized theory of gravity \cite{Weinberg65,Grishchuk77,Zeldovich86}
and which are thoroughly studied in relation with the present
development of gravitational wave detectors \cite{Schutz99,Maggiore00,Ungarelli00,Grishchuk01}.. 

Gravitational waves correspond to perturbations of the metric field and can be written in the 
transverse traceless (TT) gauge 
\beq
\label{metric_perturbation}
&&g_{\mu\nu} = \eta_{\mu\nu} + h_{\mu\nu}, \qquad \eta_{\mu\nu} = {\rm{diag}}(1,-1,-1,-1)\nonumber\\
&&h_{00} = h_{\rm i 0} = h_{\rm i}^{\rm i} = 0
\eeq
${\rm i} =1,2,3$ stands for the spatial indices whereas $0$ will represent the temporal 
index; the spatial components $h_{\mu\nu}$ of the metric tensor 
are directly connected to the Riemann curvature.
Gravitational waves are conveniently described through a
mode decomposition in space-time (coordinates $(x^\mu), x^0\equiv c t$)
\beq
h_{\mu\nu}\left( x\right) = \int \frac{{\rm d} ^4 k}{\left( 2\pi \right) ^4}
\ h_{\mu\nu}\left[ k\right] e ^{ -ik_\lambda x^\lambda }, \qquad
h_{\mu\nu} \left[ k\right] = \Sigma _\pm 
\left( \frac {\varepsilon _{\mu}^{\pm} \varepsilon _{\nu}^{\pm}} {\sqrt{2}}
\right) ^{*} h^{\pm}\left[ k\right] 
\eeq
Each Fourier component is a sum over the two circular polarizations $h^\pm$, which 
are obtained as products of the polarization 
vectors $\varepsilon^{\pm}$ well-known from electromagnetic theory.
Gravitational waves correspond to wavevectors $k$ lying on the light cone
($k^2= k_\mu k^\mu=0$), they are transverse with respect to this wavevector
($k^{\mu} \varepsilon _{\mu}^{\pm} =0$) and the metric perturbation
has a null trace ($(\varepsilon^\pm)^2 = 0$).

We consider for simplicity the case of stationary, unpolarized and isotropic 
backgounds. Then, a given metric component, say $h \equiv h_{12}$,
is a stochastic variable characterized by a noise spectrum $S_{h}$ 
\beq
\label{gw_correlation}
&&\left\langle h \left( t \right) h \left( 0 \right) \right\rangle 
= \int \frac{{\rm d} \omega}{2\pi}
\ S_{h}\left[ \omega \right] e ^{ -i\omega t }  
\eeq
$S_{h}$ is the spectral density of strain fluctuations considered in most
papers on gravitational wave detectors (see for example \cite{Maggiore00}). 
It can be written in terms of the mean number $n _\gw $ of gravitons per mode
or, equivalently, of a noise temperature $T _{\gw}$
with $\kB$ the Boltzmann constant and $G$ the Newton constant
\beq
&& S_{h} = \frac{16 G}{5 c^5} \ \hbar \omega n _{\gw}= \frac{16 G}{5 c^5} \ \kB T _{\gw} 
\label{defTgw}
\eeq 

Knowledge on gravitational wave backgrounds comes from studies  estimating 
the probability of events which might be observed by interferometric detectors of gravitational 
waves. An important component is 
 constituted by the `binary confusion background', that is
the estimated level for the background of gravitational waves emitted by unresolved 
binary systems in the galaxy and its vicinity.
This `binary confusion background' leads to a nearly flat function $S_{h}$,
that is also to a nearly thermal spectrum, in the $\mu\Hz$ to 10mHz frequency range \cite{Schutz99}
\beq
\label{gw_temperature}
10^{-6} \Hz <  \frac{\omega} {2\pi} < 10^{-4}\Hz \quad && \quad
S_{h} \sim 10^{-34} \Hz^{-1}
\eeq
With the conversion factors given above, this corresponds 
to an extremely large equivalent noise temperature $T _{\gw} \simeq 10^{41}\ \K$.
It is worth stressing that this is only an effective noise 
temperature. Such a value, larger than Planck temperature ($\sim 10^{32}\ \K$), does not correspond to 
an equilibrium temperature and is allowed by the weakness of gravitational coupling.

Previous estimations correspond to the confusion background 
of gravitational waves emitted by binary systems in our Galaxy or its
vicinity. Because of the large number of unresolved and independent sources, and
 as a consequence of the central limit theorem, they lead to a stochastic noise   
obeying gaussian statistics. 
There also exist predictions for gravitational backgrounds associated 
with a variety of cosmic processes \cite{Maggiore00}, which are however model dependent
 and have a more speculative character. 
Associated temperatures vary rapidly with frequency and are dominated by the confusion binary background 
in the frequency range considered here. 

\section{Quantum decoherence of atomic interferometers}	

Atoms used in interferometry appear as particularly interesting microscopic systems
for studying quantum decoherence, as it has recently been suggested that matter-wave interferometers could reveal the existence of intrinsic spacetime fluctuations, through an induced Brownian motion 
 \cite{Power00,Amelino00}.
Although it has not been possible to observe such an effect
in existing matter-wave interferometers, instruments are
now being designed, like the atomic interferometer HYPER for
measuring the Lense-Thirring effect in space, which possess a very high
sensitivity to gravitation fields \cite{Hyper00}. 
It is thus important, in order to confirm the viability of such instruments,
to obtain quantitative estimates of potential decoherence effects, in particular those associated with spacetime
or gravitation fluctuations.

We shall consider the atomic field of the matter-wave interferometer HYPER as a typical
example of a microscopic system affected by quantum decoherence (see for instance \cite{Borde00,Borde01,Peters01}
for details on atomic interferometry ).
 HYPER is an interferometer with a rhombic geometry which is used as a gyrometer, that is to say,
 its rotation with respect to inertial
frames is measured through the observation of a Sagnac effect.
The Sagnac dephasing $\Phi$ is  proportional 
to the mass $m_\at$ of the (non relativistic) atoms, to the area $A$ of the interferometer
and to the rotation frequency $\Omega$ 
\beq
&&\Phi ={1\over \hbar}\oint p_{\rm i}d x^{\rm i} = \frac{ 2 m_\at A} {\hbar} \Omega,
\qquad  p_\mu  = g_{\mu\nu} m_\at v^\nu_\at, \qquad  A  = v_\at^2 \tau_\at^2 \sin\alpha
\eeq
$g_{\mu\nu}$ is the metric field in the frame of the rotating interferometer,
$v_\at$ is the atomic velocity, and the area $A$ is given by the length $v_\at \tau$  of the rhomb side and the
aperture angle $\alpha$ ($\tau_\at$ is the time of flight on one rhomb side).

According to general relativity, a local inertial frame in the neighborhood of a rotating massive body 
differs from the celestial frame determined by the `fixed stars' as a consequence of the dragging 
of inertial frames. 
This gravitomagnetic (Lense-Thirring) effect in the Earth neighborhood
is measured by HYPER interferometer, by comparing the
local inertial measurement performed by the atoms to the indication of a star tracker. 
A map of the  Lense-Thirring effect around the Earth is obtained by recording the dephasings and building the 
corresponding interferogram for each position of the satellite on its orbit.

Gravitational waves, like other gravitational perturbations such as 
the Lense-Thirring effect, induce a dephasing of the matter waves within the 
two arms of the interferometer and thus affect the interference fringes \cite{Linet76}
\beq
&&\delta \Phi_\gw = {m_\at\over 2\hbar}\oint h_{\rm i j} v_\at^{\rm i} v_\at^{\rm j} d\tau
= \frac{2m_\at A} {\hbar} \delta \Omega_\gw 
\label{Phigw}
\eeq
Metric components are evaluated in the TT (transverse traceless) gauge.
Using the symmetry of the rhomb,  this expression may be 
obtained from the derivative of the metric component $h_{\rm 12}$ lying in the spatial plane defined
by the interferometer
\beq
&&\delta\Omega _\gw (t) = -\frac 12 \frac{\d \overline{h_{\rm 12}}}{\d t}, \qquad
\overline{h_{\rm 12}}(t) = \int \ h_{\rm 12} \left(t - \tau \right) 
g\left(\tau\right) \d \tau
\eeq
The linear filtering function $g$ has
a triangular shape which reflects the distribution of the time of exposition 
of atoms to gravitational waves inside the rhombic interferometer.
The square of its Fourier transform, which describes linear filtering in frequency space, 
is an apparatus function  characterizing the interferometer \cite{Lamine02}
\beq
&&|\tilde{g} \left[ \omega \right]|^2 = 
\left( \frac { \sin \frac{\omega \tau_\at}{2} } 
{ \frac{\omega \tau_\at}{2} } \right) ^4 
\eeq

We now consider the degradation of fringe contrast
obtained by averaging over stochastic dephasings. 
This evaluation \cite{Lamine02} can be shown to be equivalent
to the other approaches to decoherence (see for example \cite{Imry90}). 
Stochastic gravitational 
waves with frequencies higher than the inverse of the averaging time
identify with the unobserved degrees of freedom which
are usually traced over in decoherence theory 
(see \cite{Raimond01} and references therein).
When $\delta \Phi_\gw$
is a gaussian stochastic variable, the degraded fringe contrast is read as 
\beq
&&\left\langle \exp \left( i\delta \Phi_\gw \right) \right\rangle =
\exp \left( -\frac{ \Delta \Phi_\gw^2 } {2} \right), \qquad
\Delta \Phi_{\gw}^2 = \left\langle \delta \Phi_\gw^2 \right\rangle 
\eeq

Using the expression of $\delta \Phi_\gw$ in terms of the averaged time derivative 
of $h_{\rm 12}$ we write the variance $\Delta \Phi_\gw^2$ as an integral over 
the noise spectrum $S_h$ (\ref{gw_correlation}).
Particularly interesting is the case of an approximately flat or thermal 
spectrum $S_h$ (\ref{defTgw}) which, as discussed in previous section, 
is approximately realized by the binary confusion background on a significant frequency range. 
With a white noise assumption, the variance is found to be 
proportional to the constant value of the noise spectrum $S_h$
\beq
\label{micro_factor}
\Delta \Phi_{\gw}^2 = \left(\frac{2m_\at v_\at^2}{\hbar} \sin\alpha\right)^2\ S_h \ 2\tau_\at
\eeq
After substitution  of the numbers corresponding to 
HYPER \cite{Hyper00}, we deduce that the decoherence
due to the scattering of gravitational waves is completely negligible
\beq
\Delta \Phi^2_\gw \sim  10^{-20}  &\ & \ll 1
\eeq

 We have  discussed here
the decoherence effect on atomic fields. In fact, it appears that the decoherence effect affecting
the laser fields, involved in the stimulated Raman processes used
for building up beam splitters and mirrors for matter waves, provides a larger contribution \cite{Lamine02}.
But this  changes neither the mechanism of quantum decoherence which has been discussed here, 
nor its incidence on the instrument sensitivity. The
phase noise induced by the scattering of gravitational waves remains
completely negligible with respect to the phase noise induced by
mechanical vibrations of the mirrors. In the real instrument, decoherence
is expected to be induced by  instrumental fluctuations rather
than by  fondamental fluctuations.

\section{Quantum decoherence of planetary systems}	

After discussing the microscopic case on the example of atomic interferometers,
we come to a case which lies at the opposite end,
as it can be considered as extremely macroscopic, namely
the planetary system built by the Moon orbiting around the Earth.
The classicality of such a system may be expected to result from the strong efficiency
of decoherence mechanisms acting on it, contrarily to the case of microscopic systems.
Indeed, as we show, gravitational waves 
lead to an extremely rapid decrease of quantum coherences for such macroscopic systems.
Moreover, although decoherence may usually be attributed to collisions
of residual gaz, to radiation pressure of solar radiation or, even, to the 
scattering of electromagnetic fluctuations in the cosmic microwave background,
we show that, in the case of planetary motions, it is 
 dominated by the scattering of stochastic gravitational waves.

The Earth and Moon constitute a binary system with a large quadrupole momentum,
so that its internal motion is highly sensitive to gravitational waves.
For the sake of simplicity, we shall describe the Earth-Moon system as a circular planetary 
orbit in the plane $x_{\rm 1}x_{\rm 2}$. The  reduced mass $m$, defined from the masses of the two bodies,
will be used, such as  
the radius $\rho$, that is the constant distance between the two masses, so that
the orbital frequency $\Omega$, the normal acceleration $a$ on the circular orbit and the tangential velocity $v$ 
 obey usual relations
\beq
\label{Kepler}
a = \rho \Omega ^2 = \frac {v ^2} \rho 
\eeq

Gravitational waves will be represented as metric perturbations $h_{\mu \nu}$,
taken in the TT gauge (\ref{metric_perturbation}), so that they will be related to Riemann curvature 
($R_{\rm 0 i 0 j} = \partial_t^2 h_{\rm i j} \equiv \ddot{h}_{\rm i j}$).
The gravitational wave perturbation on the relative position $x^{\rm i}$ in the binary system 
amounts to a tidal force $\delta F$
which may also be seen as a geodesic deviation 
\beq
\label{gw_force}
&&\delta \dot{p}_{\rm i}(t) = \delta F_{\rm i}(t) = m c^2 R_{\rm 0 i 0 j} x^{\rm j}(t)
\eeq
The stochastic background of gravitational waves then induces
 a Brownian motion on the relative position of the Moon,
which  may be characterized by a momentum diffusion 
with a variance varying linearly 
with the time of exposition $\tau$ 
\beq
\label{Brownian}
&&<\delta p^2(t)> = 2 D _\gw \tau 
\eeq
The momentum diffusion coefficient $D_\gw$ is determined by the correlation function of
gravitational waves (\ref{gw_correlation},\ref{defTgw}) \cite{Reynaud01}
\beq
\label{gw_damping}
&&D_\gw = m \Gamma _\gw \kB T_\gw, \qquad \Gamma _\gw =\frac{32Gma^2}{5c^5}
\eeq
$T_\gw$ is the effective noise temperature of the gravitational background,
evaluated at twice the orbital frequency, and 
$\Gamma _\gw$ is the damping rate associated with the emission of 
gravitational waves. 
One recovers with equations (\ref{gw_damping}) 
the fluctuation-dissipation 
relation on Brownian motion \cite{Einstein05} and the quadrupole
formula for gravitational wave emission \cite{Einstein18} determined by Einstein.
Although gravitational damping can be observed in the case of
 strongly bound binary systems \cite{Taylor92},
it appears to be 
extremely small for the Moon ($\Gamma _{\gw} \approx 10^{-34}\ \s^{-1}$), with a negligible impact on its mean motion.
Moreover, it can be seen to be much smaller than the damping due to other 
environmental fluctutations, such as electromagnetic radiation pressure or
Earth-Moon tides. The latter appear to give the dominant contribution
to damping \cite{Bois96}
\beq
\label{Moon_damping}
&&\Gamma _{\gw} \ \ll \ \Gamma _{\rm em} \ < \ \Gamma _{\rm tides}
\eeq
However, as we show now, decoherence processes do not follow the same hierarchy.

Quantum decoherence may be evaluated by considering two neighbouring 
internal motions of the planetary system which correspond to the same
spatial geometry but slightly different values of the epoch, the time of passage at a given space point.
For simplicity, we measure this difference by the spatial distance $\Delta x$ 
between the two motions, which is constant for uniform motion.
The variation of momentum (\ref{gw_force}) results in a perturbation of the quantum phase 
one may associate with the relative position in the binary system
\beq
\delta\Phi_\gw(t)  = {\delta p_{\rm i}(t)\over\hbar}\Delta x^{\rm i}
\eeq
The difference of phase between two neighboring motions then undergoes a Brownian motion \cite{Reynaud01},
resulting in a random exponential factor $e ^{i \delta \Phi_\gw}$.
Averaging this quantity over the stochastic effect of 
gravitational waves, still supposed to obey gaussian statistics, 
one obtains a decoherence factor 
\beq
&&\left\langle e ^{i \delta \Phi_\gw} \right\rangle 
= \exp \left( -\frac {\Delta \Phi_\gw^2} 2 \right) 
\eeq
The decoherence factor may be expressed in terms of the variables characterizing
the Brownian motion (\ref{Brownian}) and the distance between the two motions $\Delta x$
\beq
\label{gw_decoherence}
&&\Delta \Phi_{\gw}^2 = \frac{2 D _\gw \Delta x^2 \tau} {\hbar ^2} 
\eeq
Relation (\ref{gw_decoherence}) agrees with the result expected from general discussions on decoherence 
\cite{Zurek81}: decoherence efficiency increases exponentially fast 
with $\tau$ and $\Delta x^2$.

Relation (\ref{gw_decoherence}) may be rewritten in terms of the gravitational waves spectrum (\ref{defTgw}) and
the geometric parameters of the binary system (\ref{Kepler})
\beq
\label{macro_factor}
\Delta \Phi_{\gw}^2 = \left(\frac{2mv^2}{\hbar} \sin\alpha\right)^2\ S_h \ 2\tau, \qquad 
\sin\alpha = \frac{\Delta x}{2\rho} &&
\eeq
$\frac{2mv^2}{\hbar} \sin\alpha$ is a frequency determined by the kinetic energy of the Moon
and $\sin\alpha$ is the aperture angle of the equivalent interferometer.
In the case of the Earth-Moon system, one finds  an extremely short decoherence time up to extremely
short distances $\Delta x$ (in the 
$10\mu$s range for $\Delta x$ of the order of the Planck length)
\beq
&&\frac{D_\gw}{\hbar^2} \approx 10^{75}\ \s^{-1}\m^{-2} 
\eeq

The gravitational contribution to decoherence appears to be much larger 
than the contributions associated with tide interactions and electromagnetic
scattering
\beq
&&D_\gw \ \gg \ D_{\rm tides} \ > \ D_{\rm em} 
\eeq
When compared with contributions to damping (\ref{Moon_damping}),
decoherence contributions obey a modified hierarchy. This results from their further dependence
on the level of noise induced by the environment and from the fact that gravitational waves constitute the environment
with the largest effective noise temperature (\ref{gw_temperature}).
 To be precise, the ratio 
$\frac{\Gamma_\gw}{\Gamma_{\rm tides}}$
of the damping constants associated with gravitational waves and
tides is of the order of $10^{-16}$, while the ratio $\frac{T_\gw}{T_{\rm tides}}$
is  of the order of $10^{38}$.
It follows that the ratio $\frac{D_\gw}{D_{\rm tides}}$
remains very large and that the gravitational contribution to
decoherence dominates the other ones.

The dominant mechanism leading to the classical behavior of very
macroscopic systems appears to be due to gravitational waves, originating either from the confusion
binary background in our galaxy or from extragalactic sources in
a larger region of the universe.
It is remarkable that the classicality and the ultimate fluctuations of very macroscopic systems 
appear to be determined by the classical gravitation 
theory which also explains their mean motion. 

\section{Gravitational quantum decoherence}	

The results obtained in the previous sections for gravitationally induced decoherence are 
reminiscent of the qualitative discussions of the Introduction.
 For microscopic probes, such as the atoms or photons involved in atomic interferometers,
decoherence is so inefficient that it can be ignored with the consequence
that quantum mechanics remains the appropriate description..
For macroscopic bodies on the contrary, such as the Moon-Earth system,
decoherence is extremely efficient with the consequence that potential quantum 
coherences between different positions can never be observed, leading to an appropriate 
purely classical description.. 

The scale arguments sketched in the Introduction may also be associated with precise expressions.  
In both the microscopic (\ref{micro_factor}) and macroscopic (\ref{macro_factor}) cases, 
the decoherence factor $e^{-{\Delta \Phi_\gw ^2 \over2}}$ induced by 
the gravitational environment takes a same form. It involves as an assential factor
the gravitational spectral density $S_{h}$ (\ref{defTgw}), which may be expressed as an effective noise temperature, putting into evidence its dependence on Planck mass $\mP$
\beq
&&S_{h} \simeq \Theta _\gw  \ \tP ^2, \qquad \tP ^2 = \frac{\hbar G}{c^5} = \left( \frac{\hbar}{\mP c^2} \right)^2, \qquad\Theta _\gw  \simeq \frac{\kB T_\gw }{\hbar} \simeq 10^{52}\s^{-1}
\eeq
$\Theta _\gw $ is the temperature of the
background measured as a frequency. Relations  (\ref{micro_factor}) and (\ref{macro_factor}) may then be rewritten 
\beq
&&\frac{\Delta \Phi_\gw ^2}{2} \simeq \left(\frac{ 2 mv^2\sin\alpha}{\mP c^2} \right)^2 \ 
\Theta _\gw \tau 
\label{DeltaPhi2}
\eeq
The ratio $\frac{m^2}{\mP^2}$ confirms the preliminary arguments of the Introduction, namely that
the Planck mass effectively plays a role in the definition of
a borderline between microscopic and macroscopic masses. 
However, other factors in the formula 
imply that the scaling argument on masses is not sufficient to obtain correct
quantitative estimates. The ratio of 
the probe velocity over light velocity, the equivalent aperture angle
$\alpha$ and  the frequency $\Theta _\gw$, measuring the gravitational noise level, 
enter the quantum decoherence time on an equal footing. In particular, 
the very large value of the gravitational noise level
implies that the transition between quantum and classical
behaviors could in principle be observed for masses smaller than Planck mass.
Another interesting feature is  that the parameter to be compared with Planck energy $\mP c^2$ is the
kinetic energy $m v^2$ of the probe rather than its mass energy $m c^2$.

Finally, formula (\ref{DeltaPhi2}) provides a valuable insight into the
way to design systems aiming at observing the quantum/classical transition induced by 
intrinsic gravitational fluctuations. 
The transition region $\Delta \Phi_\gw^2 \sim 1$ seems to be best approached by 
using heavy and fast  particles in a matter-wave interferometer. 
At present, interference patterns have been observed on rather large molecules
\cite{Hornberger03,Hacker04}. But one checks that, in these experiments, the kinetic energy of the molecules, 
the area and aperture angle of the interferometer are such that the 
gravitational quantum decoherence remains negligible,
as in HYPER. Increasing these sensitive parameters so that the transition
could be approached appears as a formidable experimental challenge \cite{Lamine06} (see
\cite{Houde00,Boustimi01} for using fast molecules).
Alternatively, one could consider using quantum condensates
\cite{vanderWal00,Li06},
an approach however requiring further technological progress.

\end{document}